\documentclass[acmtog]{acmart}

\AtBeginDocument{%
  \providecommand\BibTeX{{%
    \normalfont B\kern-0.5em{\scshape i\kern-0.25em b}\kern-0.8em\TeX}}}

\setcopyright{acmcopyright}
\copyrightyear{2020}
\acmYear{2020}
\acmDOI{10.1145/1122445.1122456}

\acmJournal{TOG}
\acmVolume{37}
\acmNumber{4}
\acmArticle{111}
\acmMonth{8}
\pdfoutput=1

\usepackage{enumitem}
\usepackage[boxruled, linesnumbered, noend, noline]{algorithm2e}
\newlist{steps}{enumerate}{1}
\setlist[steps, 1]{label = Step \arabic*:}
\usepackage{listings}
\usepackage{graphicx}
\usepackage{latexsym}
\usepackage{etoolbox}
\usepackage{academicons}

\definecolor{codegreen}{rgb}{0,0.6,0}
\definecolor{codegray}{rgb}{0.5,0.5,0.5}
\definecolor{codepurple}{rgb}{0.58,0,0.82}
\definecolor{backcolour}{rgb}{0.95,0.95,0.92}

\lstdefinestyle{mystyle}{
  backgroundcolor=\color{backcolour},   commentstyle=\color{codegreen},
  keywordstyle=\color{magenta},
  numberstyle=\tiny\color{codegray},
  stringstyle=\color{codepurple},
  basicstyle=\ttfamily\footnotesize,
  breakatwhitespace=false,         
  breaklines=true,                 
  captionpos=b,                    
  keepspaces=true,                 
  numbers=left,                    
  numbersep=5pt,                  
  showspaces=false,                
  showstringspaces=false,
  showtabs=false,                  
  tabsize=2
}

\date{ }

\begin{document}

\title{Huskysort}

\author{R C HILLYARD}
\email{r.hillyard@northeastern.edu}
\authornote{
    originator of the idea; did the analysis; designed, coded, and performed much of the benchmarking.
    The developed Huskysort Code is available at: {\url{https://github.com/rchillyard/HuskySort}}
    }
\affiliation{
  \institution{College of Engineering, Northeastern University}
  \streetaddress{360 Huntington Avenue}
  \city{Boston}
  \state{Massachusetts}
  \country{USA}
  \postcode{02115}
}

\author{Yunlu Liaozheng}
\email{liaozheng.y@northeastern.edu}
\authornote{
    implemented the initial algorithm, including the coders and the key innovation of the object swap in the quicksort phase; early benchmarking.
    }
\affiliation{
  \institution{College of Engineering, Northeastern University}
  \streetaddress{360 Huntington Avenue}
  \city{Boston}
  \state{Massachusetts}
  \country{USA}
  \postcode{02115}
}

\author{Sai Vineeth K R}
 \email{kandappareddigari.s@northeastern.edu}
 \orcid{0000-0001-9734-7358}
 \authornote{Analysed the data, extended the literature survey and wrote the paper with input from all authors.}
\affiliation{
  \institution{College of Engineering, Northeastern University}
  \streetaddress{360 Huntington Avenue}
  \city{Boston}
  \state{Massachusetts}
  \country{USA}
  \postcode{02115}
  \orcid{0000-0001-9734-7358}
}



\begin{abstract}
Much of the copious literature on the subject of sorting has concentrated on minimizing the number of comparisons and/or exchanges/copies.
However, a more appropriate yardstick for the performance of sorting algorithms is based on the total number of array accesses that are required (the "work").
For a sort that is based on divide-and-conquer (including iterative variations on that theme), we can divide the work into linear, i.e. $\textbf{O}(N)$, work and linearithmic, i.e. $\textbf{O}(N log N)$, work.
An algorithm that moves work from the linearithmic phase to the linear phase may be able to reduce the total number of array accesses and, indirectly, processing time.
This paper describes an approach to sorting which reduces the number of \emph{expensive} comparisons in the linearithmic phase as much as possible by substituting inexpensive comparisons.
In Java, the two system sorts are dual-pivot quicksort (for primitives) and Timsort for objects.
We demonstrate that a combination of these two algorithms can run significantly faster than either algorithm alone for the types of objects which are expensive to compare.
We call this improved sorting algorithm Huskysort.
\end{abstract}



\maketitle

\section{Introduction}
With the vast increase in the size of the data world, efficient computing algorithms play a crucial role in processing data.
From the earliest days of computing, sorting has been one of the most studied areas of research even though it appears to be a straightforward and familiar topic.
The rise of data volume and complexity has made clear a need for new approaches to traditional sorting techniques.
Within the realm of comparison-sort algorithms, many aspects of the algorithm can be significant: stability, extra memory, partitioning, merging, exchanging (swapping), copying, and, of course, the comparison itself.

Even though sorting appears to be a settled area of research, yet new improvements are still being made: \cite{Dual-pivot,Timsort,doi:10.1137/1.9781611975994.101,10.5555/313559.313859}.
There is, therefore, an abundance of sorting techniques available.
Nevertheless, there is still room for improved versions in regard to computation, memory, time, and space complexity.
In a recent paper \cite{sort}, the authors conclude that quicksort is indeed the fastest general-purpose sorting algorithm, especially when appropriate care is taken to avoid the quadratic worst-case with several other improvements (e.g. dual-pivot quicksort)\cite{Dual-pivot} over the traditional version \cite{10.1145/366622.366644}.
However, quicksort does not perform as well as some variations of merge sort, especially Timsort \cite{Timsort}, when the dataset is partially sorted.
Timsort, a hybrid stable sorting algorithm derived from merge sort, is now available in the Python and Java libraries as the default object-sorting utility.

We determined to test the efficacy of combining quicksort with Timsort for object types that are costly to compare.
Such types include \emph{String, LocalDateTime, BigInteger, BigDecimal}, and many user-defined types of this kind.

We review existing methods of sorting in Section 2 and go through the Huskysort algorithm in Section 3.
Implementation options and their likely impact are explored in Section 4.
We describe our Test Cases in Section 5 and present results in Section 6, with outcomes summarized in Section 7.

\section{BACKGROUND}
Timsort is ideally suited to the partially ordered data which is common in the real-world business applications.
For this use case it is easily the fastest standard algorithm.
It is less well suited, however, for another use case, which is unordered data.
Huskysort is designed specifically for this use case: unordered data consisting of objects which are relatively expensive to compare.

Before talking about our approach let's look at insertion sort and Timsort in more detail.
The key feature of both of these methods is that the number of comparisons (and swaps, in the case of insertion sort) is proportional to the number of inversions.
Because insertion sort is an "elementary" sort (it reduces the complexity of the problem by iteratively partitioning by one element at a time), it is only suitable for small arrays.
It traverses the array, moving each new element encountered into its proper place in the ordered partition, which therefore grows by one element in each iteration until the whole array is ordered.

Timsort is a variation of bottom-up merge sort which takes advantage of existing ordered runs.
Because of this, it is very efficient for partially-ordered arrays.
Additionally, like merge sort, Timsort is stable.
Timsort's worst case is as $\textbf{O}(N + N log p)$, where $p$ is the number of runs \cite{DBLP:journals/corr/abs-1805-08612}.

One of the authors (Hillyard) had set a question in a mid-term exam (Fall, 2018) about a hypothetical algorithm that created a 64-bit hash code for a date-time object and then used quicksort to sort the array.
Afterward, it became apparent that it was not such a crazy idea after all.
A quick implementation showed that, indeed, we could see an improvement over the system sort (Timsort) for sorting objects.
The essence of the algorithm lies in the fact that quicksort can sort an array of (64-bit) longs very quickly.
64 bits affords sufficient room to encode many objects uniquely, but even if the coding is not unique, post-sorting an array of objects which are quasi-ordered can be very fast using Timsort.
The key innovation of the algorithm was developed by Liaozheng: swap the objects at the same time as swapping the longs.

\section{ALGORITHM}

\subsection{Overview}

\begin{figure}[]
  \caption{Huskysort Flow}
  \label{Fig:HS}
   \Description{An Overview of Huskysort}
  \centering
  \includegraphics[width=0.5\textwidth]{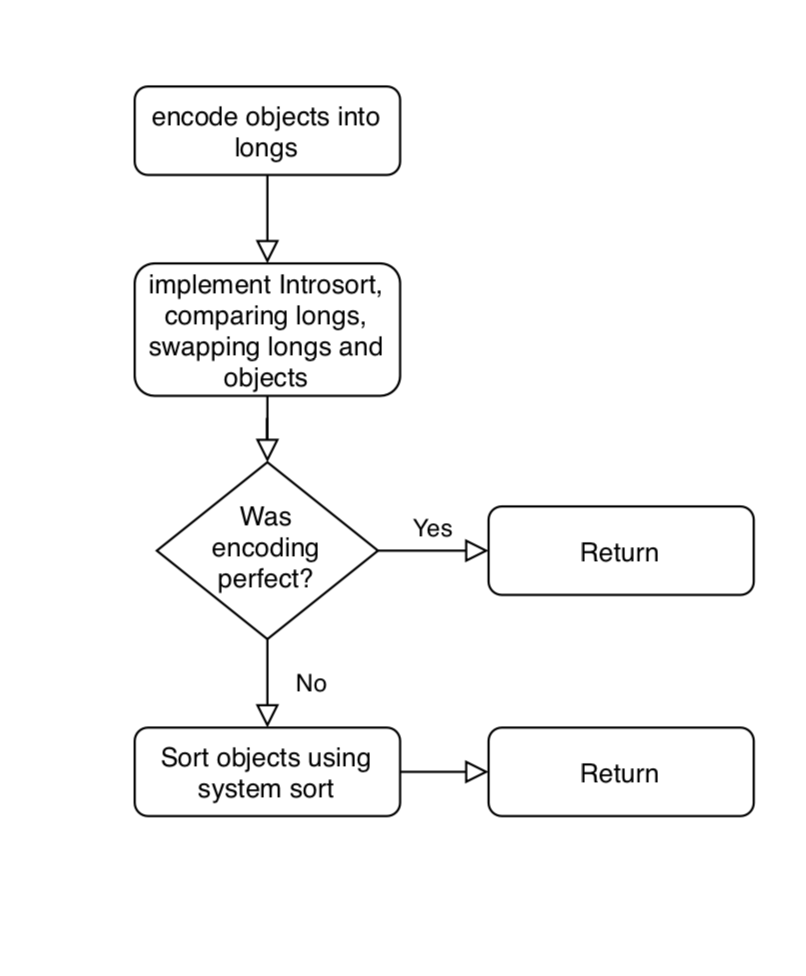}
\end{figure}
\begin{figure}[]
  
  \caption{Introsort Flow}
  \label{Fig:Introsort}
  \Description{An Overview of Introsort}
  \centering
  \includegraphics[width=0.5\textwidth]{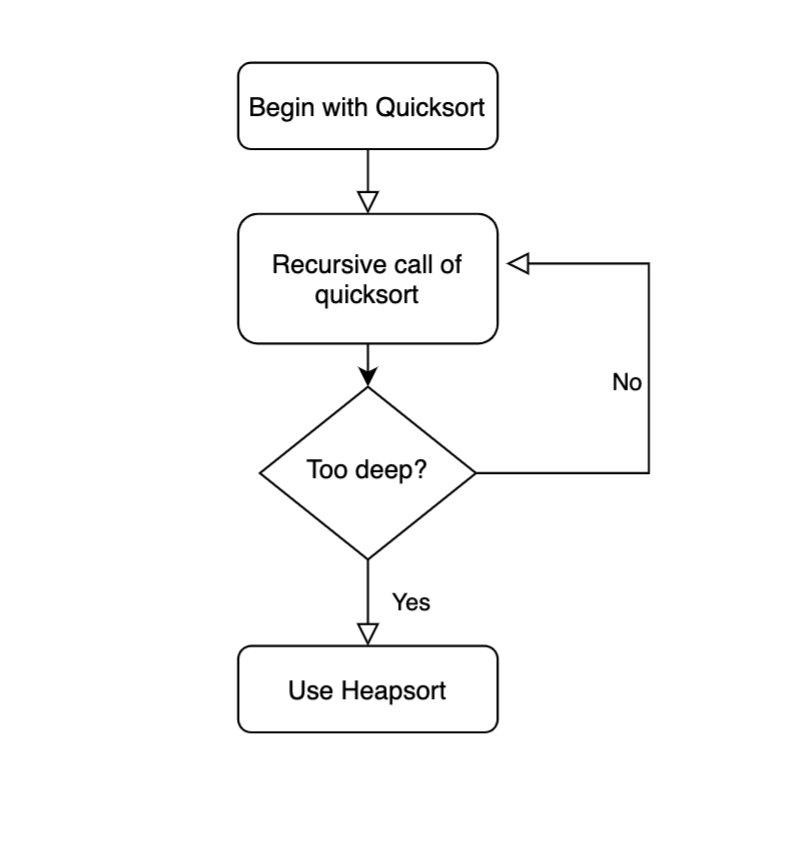}
\end{figure}
The strategy of Huskysort (see Flow ~\ref{Fig:HS}) is as follows (assuming that the dataset is presented as an array $xs$ of type $X$): see Algorithm ~\ref{alg:the_alg}

\begin{steps}
\item  
\renewcommand{\labelenumi}{\alph{enumi}}
    \begin{enumerate}
    \item Create a $long[]$ array (\emph{longs}) of the same size as $xs$;
    \item for each element $x$ of $xs$, set the corresponding element of \emph{longs} to $x.huskyCode()$.
    \end{enumerate}
\item {}Sort array \emph{longs} using quicksort (in practice, we have found that
Introsort \cite{doi:10.1002} works well) in such a way that when elements $i, j$ of \emph{longs} are
exchanged, then elements $i, j$ of $xs$ are also exchanged; See Flow ~\ref{Fig:Introsort}
\item {}Finally, when quicksort is finished and, if necessary, we run Timsort on the array $xs$ to ensure that its elements are truly in order.
\end{steps}

For the details of the husky-coding itself, see Algorithm ~\ref{alg:the_alg4}

\subsection{Discussion of Husky Encoding}

Step 1b requires the definition of a 64-bit hash function (huskyCode) whose properties are as follows: (refer to Algorithm ~\ref{alg:the_alg2})
\begin{itemize}
                        \item {}If $x = y$, then $x.huskyCode() = y.huskyCode()$;
                        \item {}If $x > y$, then: 
                            \begin{itemize}
                            \item{} either $x.huskyCode() <= y.huskyCode()$ (with probability $p$)
                            \item{} or $x.huskyCode() > y.huskyCode()$ (with probability $1-p$);
                            
                            \end{itemize}
                        
                        \end{itemize}

We interpret this rule as follows: if $x$ and $y$ are the same ($x.equals(y)$ in Java), their huskyCode should be the same.
If $x > y$, we normally expect that $x.huskyCode() > y.huskyCode()$, i.e. the sort on the husky-coded longs will not invert $x$ and $y$.
However, when $x > y$ while $x.huskyCode() < y.huskyCode()$, then an inversion will be introduced.
And, when $x > y$ and $x.huskyCode() = y.huskyCode()$, then an inversion \emph{may} be introduced because quick-sort is not stable.\cite{6115063}

For a successful encoding, the value of $p$, the probability of a potential inversion, should be less than some critical probability, $p_{crit}$.
Leaving a more detailed discussion of $p_{crit}$ until later, we note that, for some domains of $X$, we can definitely assert that $p = 0$.
We refer to this as a “perfect” encoding.
In such a case, we can skip step 3 of the algorithm because, when $p = 0$, there will be no inversions remaining after quick-sorting according to the huskyCodes.
An example of such a domain is date-time values with a resolution of one nano-second over a period of 128 years.
For strings of English case-independent letters (no punctuation or numbers), i.e. 5-bit ASCII, words of 12 (or fewer) characters in length can also be coded uniquely.
For case-dependent strings (6-bit ASCII), we can encode perfectly if the lengths are all 10 characters or less.
In step 3, it might be acceptable to use insertion sort instead of Timsort.
In any event, assuming that the huskyCode is as accurate as required, then there will be very few inversions (elements out of place) remaining after the quicksort phase.
This implies that the second sort can be performed in very close to linear time.
The results have been very satisfactory.
For example, taking an array of 100,000 English words chosen randomly from a corpus of 100,000 words \cite{GOLDHAHN12.327}, we have found, for example in one run, that Huskysort takes a normalized time of 0.93, compared with 1.00 for quicksort and 1.45 for Timsort (For further data, please refer to table ~\ref{tab:HSComp}.)

\subsection{Discussion of $p_{crit}$}
When $p > 0$, there is a finite probability of any pair of elements remaining inverted after the first sorting pass.
For an array of N randomly ordered elements, the average number of inversions before sorting is $X = N (N -1 ) / 4$.
The number of inversions remaining after the first pass is simply $p X$.
And, for Timsort or insertion sort, the time to re-sort the array will be $t = k (N + p X)$ where $k$ is some system-dependent constant.
The total time to sort the array is made up of these three time-consuming steps where $k_1, k_2, k_3$ are system- and algorithm-dependent constants:

\begin{steps}
 \item{} $T_1$ is the time to encode the elements:
\[T_1 = k_1 N\]
\item{}  $T_2$ is the (average) time to quick-sort the huskyCode array while also swapping the element array itself:
\[T_2 = 2 k_2 N ln N\]
\item{}  $T_3$  is the time to Timsort the element array (only required if $p>0$):
 \[T_3 = k_3 (N + p X)\]
\end{steps}
Provided that the total time $T = T_1 + T_2 + T_3$ is less than the time required to use Timsort on the original element array, we have a win.
The actual values of $p_{crit}$ are very much dependent on the machine running the sort.
We have not calculated a typical value for $p_{crit}$ but, even for the worst-case: Chinese characters using UTF8 encoding, we still find very significant gains.
Please refer to table ~\ref{tab:Improvements Summary}	

\subsection{Explanation of Working}

So, why does this work? Well, first of all, we observe that 64-bit word length in today’s computers is the norm.
This allows for an encoding which has a low, possibly zero, value of $p_{crit}$.

Even though a Unicode string can encode only the first 3 15/16ths characters in 64 bits, this is extremely helpful in getting the elements close to their final order.

Here, we may compare the method of Huskysort with that of ShellSort.\cite{10.1007/3-540-61680-2_42}

ShellSort acts exactly like a series of insertion sorts but, whereas each compare/swap in insertion sort is limited to resolving one inversion only, so in ShellSort, $h$ inversions can be resolved in one compare/swap (where we are $h$-sorting the array).
So, Shellsort’s primary mechanism is to try to fix as many inversions as possible before taking the subsequent step.

In the same way, Huskysort fixes as many inversions as possible before doing the final step.
Note that it isn’t necessary for the encoding to be perfect.
But it should be good.

Another way of looking at it is that running quicksort or merge sort (or their variations) will require $\textbf{O}( N log N )$ comparisons.
Merge sort, like insertion sort, is linear when the number of inversions is small and grows with $\textbf{O}(N)$.
But merge sort does use extra memory.
But values such as 64-bit integers (int or long), doubles, can be compared extremely quickly in modern machines—far quicker than the time required to follow a reference (pointer) to an object in memory.
In Java, this distinction is described by contrasting primitives with objects.
Primitives can be quick-sorted very fast and of course, long is primitive.

Exchanging (swapping), on the other hand, takes the same amount of time whatever type you are sorting because swapping an object reference is just the same as swapping any 64-bit word.
The other reason why quicksort is so fast is that it is cache-friendly.
So, we perform a fast quicksort on the array of longs.
The only difference between the Husky version of quicksort and standard quicksort is that the swap operation must actually do two swaps: one of the longs and one of the object references.
But, again, swapping is fast because elements of both arrays (i.e. both longs and refs) have good chances of being cached.

The extra time (and space) required to perform the encoding upfront, which is \textbf{O}(N), is more than offset by the faster comparisons of longs as opposed to comparisons of objects.

After the first pass (using quicksort), we can easily determine whether our encoding was perfect—in which case, there will be no remaining inversions.
We do this for non-sequence classes via a simple constant-time lookup.
For sequences (in practice, these are always Strings), we check each sequence’s coding perfection (according to its length), and as soon as we find a sequence that cannot be encoding perfectly, we stop checking.

If the coding was perfect, we simply skip the final pass.

Given the notion of hashing that the algorithm uses, it would seem natural to call is Hash Sort.
Unfortunately, that name was already taken for quite a different algorithm \cite{DBLP:journals/corr/cs-DS-0408040}.

\begin{algorithm}
\KwIn{\it{xs} as an array of object X which extends Comparable<X>}
 coding = huskyCoder.huskyEncode(xs)\;
 longs = coding.longs\;
 Introsort(xs, longs, 0, longs.length, 2 * floor\_lg(xs.length))\;
 \If{coding.perfect}{return\;}
 Arrays.sort(xs)\;
 \caption{Huskysort.sort}
 \label{alg:the_alg}
\end{algorithm}

\begin{algorithm}
\KwIn{\it{xs} as an array of object X}
\KwOut{A new \it{Coding} object}
 result = new long[xs.length]\;
 \For{i = 0 to xs.length}{
 result[i] = huskyEncode(xs[i])\;
 }
 return new Coding(result, perfect())\;
 \caption{huskyEncode}
 \label{alg:the_alg2}
\end{algorithm}

\begin{algorithm}
\KwIn{\it{xs} as an array of CharSequence X}
\KwOut{A new \it{Coding} object}
 isPerfect = true\;
 result = new long[xs.length]\;
 \For{i = 0 to xs.length}{
 X x = xs[i]\;
 \If{isPerfect}{isPerfect = perfectForLength(x.length())\;}
 result[i] = huskyEncode(xs[i])\;
 }
 return new Coding(result, isPerfect)\;
 \caption{huskyEncode (for character sequence)}
 \label{alg:the_alg3}
\end{algorithm}

\begin{algorithm}
\KwIn{\\
\it{str} as String\; 
\it{maxLength} as the max length can be encoded\;
\it{bitWidth} as how many bits will one char be encoded into\;
\it{mask} as a modifier for different encoding\;
}
\KwOut{A new \it{Coding} object}
 length = Math.min(str.length(), maxLength)\;
 padding = maxLength - length\;
 result = 0\;
 \eIf{mask == 0}{
 \For{i = 0 to length}{
 result = result << bitWidth | str.charAt(i)\;
 }
 }{
 \For{i = 0 to length}{
 result = result << bitWidth | str.charAt(i) {\&} mask\;
 }
 }
 result = result << bitWidth * padding\;
return result\;
 \caption{stringToLong}
 \label{alg:the_alg4}
\end{algorithm}

\section{Implementation}

\subsection{Data Source}
The sources for the Strings (English and  Chinese) are retrieved from Leipzig Corpora Collection \cite{GOLDHAHN12.327}.

In each case, we take data from the appropriate \emph{sentences.txt} file and break it up into strings as shown in  code~\ref{lst:DS}

\begin{figure}[]
  \caption{Strings Data Example}
  \label{fig:Example}
   \Description{An Example of Data file}
  \centering
  \includegraphics[width=0.5\textwidth]{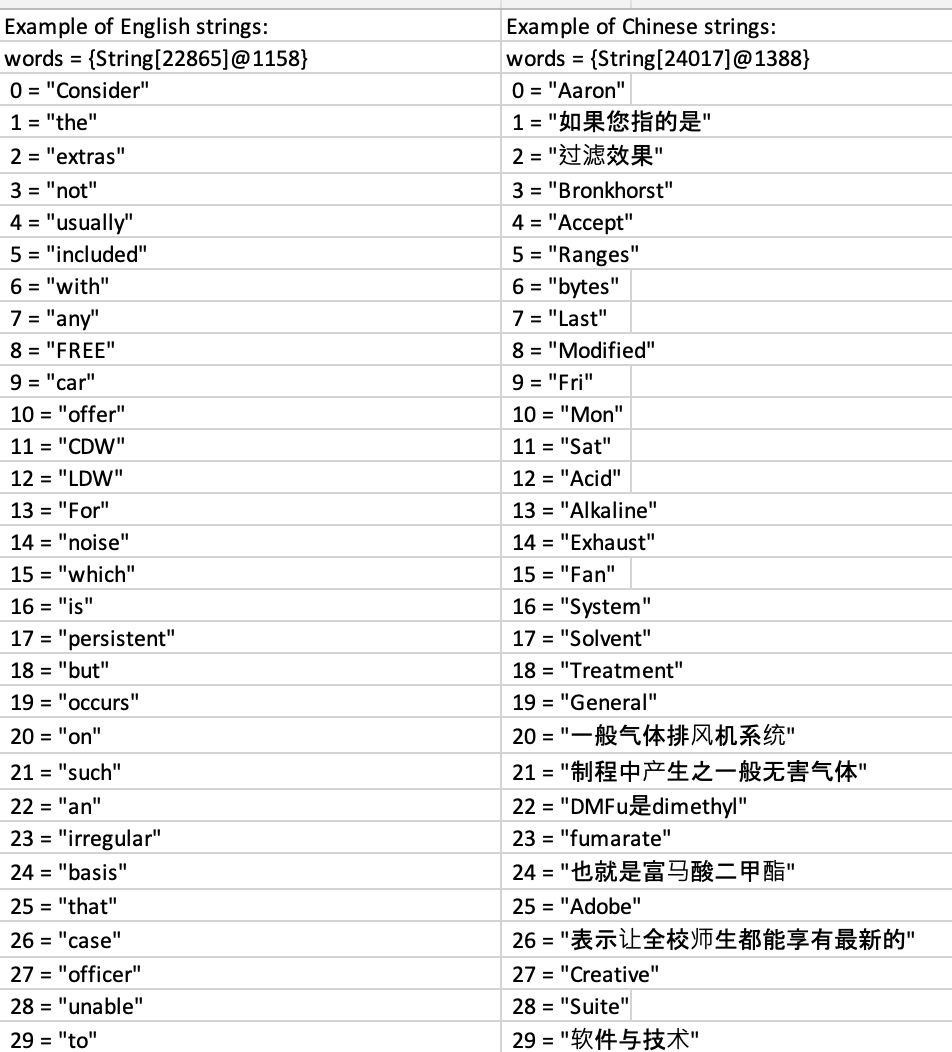}
\end{figure}

The sample of data can be seen here ~\ref{fig:Example}

\begin{lstlisting}[caption={Data Source},label={lst:DS},language=Java]
private static String[] getLeipzigWordsFromResource(final String resource) {
    return getWords(resource, HuskySortBenchmark::getLeipzigWords);
}
private static List<String> getLeipzigWords(final String line) {
    return HuskySortBenchmarkHelper.splitLineIntoStrings(line, REGEX_LEIPZIG, REGEX_STRINGSPLITTER);
}
final static Pattern REGEX_LEIPZIG = Pattern.compile("[~\\t]*\\t(([\\s\\p{Punct}\\uFF0C]*\\p{L}+)*)");
public static final Pattern REGEX_STRINGSPLITTER = Pattern.compile("[\\s\\p{Punct}\\uFF0C]");


static List<String> splitLineIntoStrings(final String line, final Pattern lineMatcher, final Pattern stringsplitter) {
    final Matcher matcher = lineMatcher.matcher(line);
    if (matcher.find()) return Arrays.asList(stringsplitter.split(matcher.group(1)));
    else return new ArrayList<>();
}

static String[] getWords(final String resource, final Function<String, List<String>> stringListFunction) {
    try {
        File file = new File(getPathname(resource, QuickHuskySort.class));
        return getWordArray(file, stringListFunction, 2);
    } catch (FileNotFoundException e) {
        logger.warn("Cannot find resource: "+resource, e);
        return new String[0];
    }
}
    } catch (final Exception e) {
        throw new RuntimeException(e);
    }
}


\end{lstlisting}

\subsection{System Environment}

\begin{figure}[]
  
  \caption{System Environment}
  \label{fig:SE}
  \Description{System configuration }
  \centering
  \includegraphics[width=0.5\textwidth]{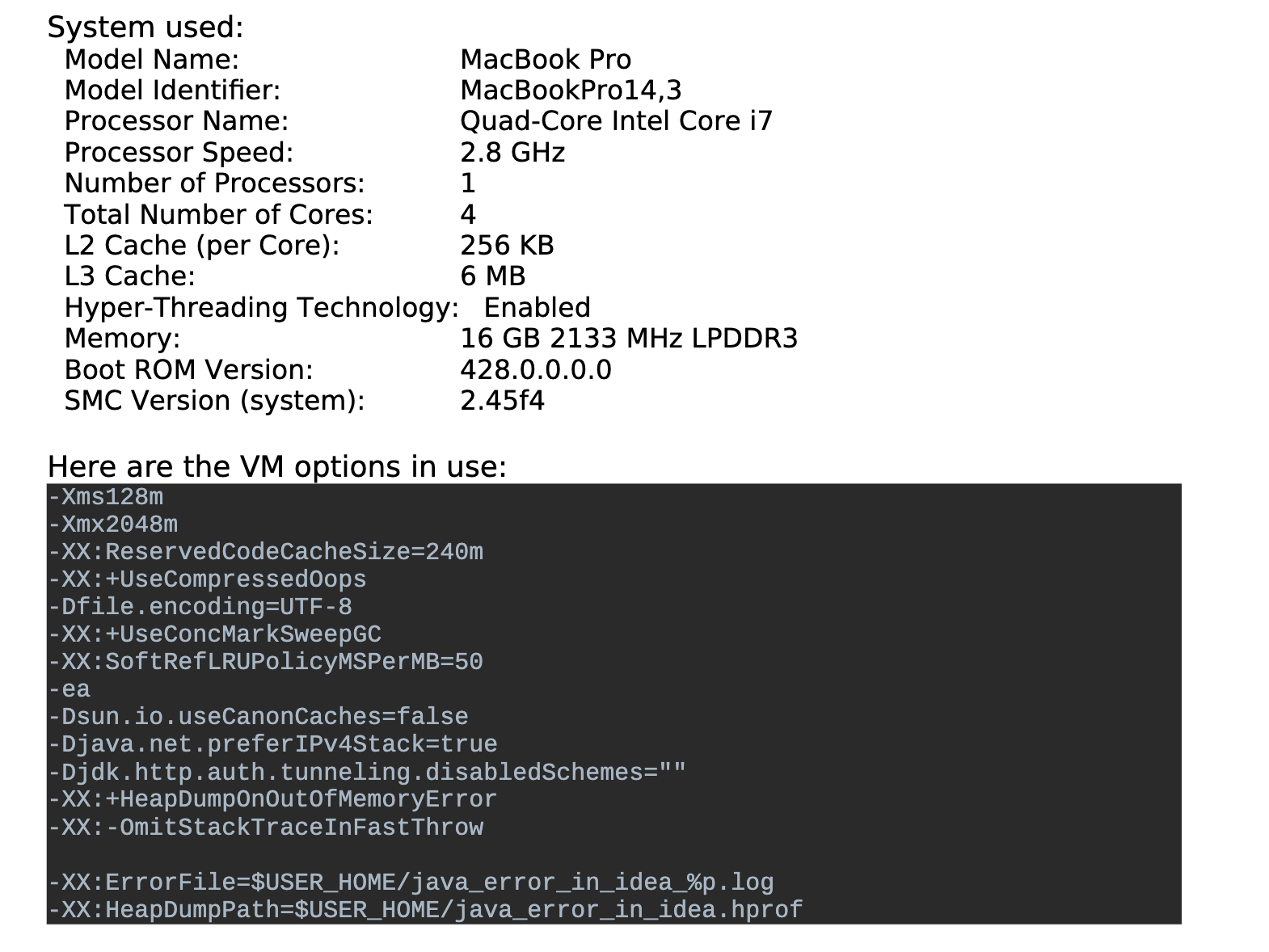}
\end{figure}

The configurations of the system that is used for implementation is shown in figure ~\ref{fig:SE}.
The Java Virtual Machine running is 1.8.0\_152.

\subsection{Implementation of Algorithm}
\begin{itemize}
                        \item {} Huskysort (top level)
\begin{lstlisting}[language=Java, caption=Husky Overview]
public void sort(final X[] xs) {
        // NOTE: we start with a random shuffle
        // This is necessary if we might be sorting a pre-sorted array. Otherwise, we usually don't need it.
        if (mayBeSorted) Collections.shuffle(Arrays.asList(xs));
        // NOTE: First pass where we code to longs and sort according to those.
        final Coding coding = huskyCoder.huskyEncode(xs);
        final long[] longs = coding.longs;
        introSort(xs, longs, 0, longs.length, 2 * floor_lg(xs.length));
        // NOTE: Second pass (if required) to fix any remaining inversions.
        if (coding.perfect)
            return;
        Arrays.sort(xs);
    }
\end{lstlisting}

                \item {} Huskysort (Husky Encoding)
                \begin{lstlisting}[language=Java, caption=HuskyEncode]
 default Coding huskyEncode(X[] xs) {
        long[] result = new long[xs.length];
        for (int i = 0; i < xs.length; i++) result[i] = huskyEncode(xs[i]);
        return new Coding(result, perfect());
    }
\end{lstlisting}

                \item {} Class to combine the long codes for an array of objects with a determination of coding perfection.
                \begin{lstlisting}[language=Java, caption=HuskyEncode]
public class Coding {
    public Coding(long[] longs, boolean perfect) {
        this.longs = longs;
        this.perfect = perfect;
    }
    public final long[] longs;
    public final boolean perfect;
}
\end{lstlisting}

                \item {} Unicode to Long :
                \begin{lstlisting}[language=Java, caption=unicodeToLong]
 private static long unicodeToLong(final String str) {
        return stringToLong(str, MAX_LENGTH_UNICODE, BIT_WIDTH_UNICODE, MASK_UNICODE) >>> 1;
    }
    \end{lstlisting}
    
             \item {} String To Long :
                \begin{lstlisting}[language=Java, caption=stringToLong]
 private static long unicodeToLong(final String str) {
        return stringToLong(str, MAX_LENGTH_UNICODE, BIT_WIDTH_UNICODE, MASK_UNICODE) >>> 1;
    }
    \end{lstlisting}
    
             \item {} Ascii to Long:
                \begin{lstlisting}[language=Java, caption=asciiToLong]
private static long asciiToLong(final String str) {
    return stringToLong(str, MAX_LENGTH_ASCII, BIT_WIDTH_ASCII, MASK_ASCII);
}
    }
\end{lstlisting}

                \item {} Constants:
                \begin{lstlisting}[language=Java, caption=Constants]
private static final int BITS_LONG = 64;
private static final int BIT_WIDTH_UNICODE = 16;
private static final int MAX_LENGTH_UNICODE = BITS_LONG / BIT_WIDTH_UNICODE;
private static final int MASK_SHORT = 0xFFFF;
private static final int MASK_UNICODE = MASK_SHORT;
private static final int BIT_WIDTH_ASCII = 7;
private static final int MAX_LENGTH_ASCII = BITS_LONG / BIT_WIDTH_ASCII;
private static final int MASK_ASCII = 0x7F;
\end{lstlisting}

\end{itemize}

\section{Test Case and Analysis}
Let’s try to do an analysis of Huskysort's performance.
The first thing to note is that, like merge sort, Huskysort requires $\textbf{O}(N)$ extra space.
This is because, in addition to the array of object references, we must have an equal-length array of longs (the huskyCode values).

Secondly, Huskysort is not stable unless using the merge sort variation.
How many comparisons and swaps does Huskysort require, on average? It has been shown for dual-pivot quicksort [Nebel] that these values are (approximately):
\begin{itemize}
    \item Comparisons: $1.9 N ln N$;
    \item Swaps: $0.6 N ln N$.
\end{itemize}

Our experiments [refer to Quicksort Analysis table] show that each comparison requires 2 array accesses (although, because one comparand is always a pivot value, this number is effectively closer to 1.
Each swap requires 4 array accesses, although again, we will be swapping elements that have already been accessed.
The effective number of (normalized) array accesses is, in total, approximately 4 rather than the expected 3.8 + 2.4.
That is to say, the coefficient of $N ln N$ is approximately 4.
The analysis represented by the table is for dual-pivot quicksort.
\footnote{Note that, since dual-pivot quicksort is not actually implemented in the Java library for Comparable objects, we re-implemented the class for objects.}
However, in Huskysort, each swap also must swap the object references.
These will not, in general, have been pre-fetched and so the effective total normalized array accesses ($A$) is approximately:
\[A = 4 + 0.6 * 4 = 6.4\]
Additionally, Huskysort requires $N$ huskyCode constructions.
Each of these requires $k+1$ array accesses where $k$ is proportional to the number of fields contributing to the huskyCode.
For a string, that will be the number of characters in the string (except that this is limited to a relatively small number according to the encoding).
For ASCII strings, it is nine.
For Unicode strings, it is four.

However, keep in mind that this is linear and, while it can’t be completely ignored, it doesn't contribute to the overall growth of the complexity.
Once the Huskysort is complete, we will assume that there are $p N$ remaining inversions, where $p$ is the probability of an idiogenic (self-inflicted) inversion.
Again, while we should not ignore it completely, it is linear.
The number of array access for the comparisons/swaps to fix these will be $(2 + j + 4) p N$, where $j$ is the number of fields to be compared to get a result ($j$ is usually smaller than $k$).

So, the total number of array accesses ($A$) for Husky-sorting an array of length $N$ is:
\[A = (6.4 ln N + (2 + j + 4) p + k + 1) N\]
   
Timsort would be hard to analyze in the general case and so we instead analyze merge sort, which should be approximately the same number of array accesses in the average case.
 
For merge sort, the analysis is much simpler: the number of comparisons for a randomly ordered array of length $N$ is $N log_2 N$.
Merge sort does not use swaps, \emph{per se}, but instead uses copies.
Assuming that we optimize the extra copying, there will be $N$ copies to get the algorithm started and $N$ per level.
Thus, the number of copies will be $N {log_2 N}$.
Each of these copies requires two array accesses while each comparison requires 2 + j (described above) array accesses.
 
The total number of array accesses ($A$) for merge sort is, therefore:
\[A = (4 + j) N log_2 N + 2 N\]
 
Given that $log_2 N = 1.44 ln N$, and assuming that:

$j = 4$ corresponding to the comparison of two strings that start with the same character but differ in the second character; and

$k = 7$ and $p = 0.1$; then

the totals ($A_m$ for merge sort and $A_h$ for Huskysort) come to:
 \[A_m = 11.5 N ln N + 2 N\]
 \[A_h = 6.4 N ln N + 9 N\]
 
The consequence of this is that, as expected, the linear phase of Huskysort is relatively high for small $N$ but that is soon swamped by the saving of comparisons in the linearithmic phase.
The break-even point is actually a very small value of $N$: four.
As $N$ increases, we expect the advantage of Huskysort to increase further.
This is generally borne out by the benchmarks.
Please refer to tables ~\ref{tab:Comparison} and ~\ref{tab:HSComp}.
 \begin{table*}
  \caption{Theoretical comparisons}
  \label{tab:Comparison}
  \begin{tabular}{cccc}
    \toprule
    N  &  N ln N & merge sort & Huskysort \\
    \midrule
     \ 4 &6 &72 &72 \\
     \ 1,000 &6,908 &81,439  &55,075  \\
     \ 1,000,000 &13,815,511  &160,878,371  &101,149,455  \\
     \ 1,000,000,000 &20,723,265,837  &240,317,557,125  &147,224,183,132  \\

    \bottomrule
  \end{tabular}
\end{table*}

\begin{table*}
  \caption{Relative sort times for Huskysort, dual-pivot quicksort, system sort}
  \label{tab:HSComp}
  \begin{tabular}{cccc}
    \toprule
   N&Huskysort&DP quicksort&system sort\\
    \midrule
    \ 1000&0.60&	1.00&0.84\\
\ 2000&0.78&	1.00&1.19\\
\ 4000&0.81&	1.00&1.23\\
\ 8000&0.77&	1.00&1.27\\
\ 16000&0.74&1.00&1.24\\
\ 32000&0.74	&1.00&1.24\\
\ 64000&0.85	&1.00&1.29\\
\ 125000&0.76&1.00&1.51\\
\ 250000&0.92&1.00&1.32\\
\ 500000&0.97&1.00&1.42\\
\ 1000000&0.95&1.00&1.45\\
\ 2000000&0.93&1.00&1.54\\
\ 4000000&0.91&1.00&1.64\\

    \bottomrule
  \end{tabular}
\end{table*}
Since the final pass of Huskysort will be cleaning up a relatively small number of inversions, does it matter whether we use insertion sort or Timsort? It turns out that for arrays which are smaller than about 50,000 Strings, it makes barely any difference at all.
Indeed, insertion sort is sometimes faster.
However, as the size of the array increases beyond this value, Timsort begins to win out over insertion sort. Please refer to table ~\ref{tab:TimvsInsertion}.

 \begin{table*}
  \caption{Timsort or insertion sort}
  \label{tab:TimvsInsertion}
  \begin{tabular}{ccc}
    \toprule
    N & Husky with Tim & Husky with Insertion\\
    \midrule
    \ 1000&0.14&0.14 \\
\ 2000&0.31&0.30 \\
\ 4000&0.63&0.64 \\
\ 8000&1.46&1.56 \\
\ 16000&3.24&3.19 \\
\ 32000&6.56&6.48 \\
\ 64000&16.85&19.41 \\
\ 125000&47.74&54.90 \\
\ 250000&168.21&167.90 \\
\ 500000&259.37&373.43 \\
\ 1000000&602.08&1202.61 \\
\ 2000000&1234.90&3129.23 \\
\ 4000000&2423.62&11018.19 \\

    \bottomrule
  \end{tabular}
\end{table*}

\section{Results and Observations}
\subsection{Benchmarks}
\begin{itemize}
\item{} Comparison of sort times for Huskysort, dual-pivot quicksort, system sort.
Please refer to table ~\ref{tab:HSComp}.

\item {}Time vs Size ($N$) for Huskysort,
Please refer to figure ~\ref{Fig:TvsN}
\begin{figure}[]
  
  \caption{Time vs Size ($N$)}
  \Description{Time taken with Huskysort Corresponding to  Size of Words}
  \label{Fig:TvsN}
  \centering
  \includegraphics[width=0.5\textwidth]{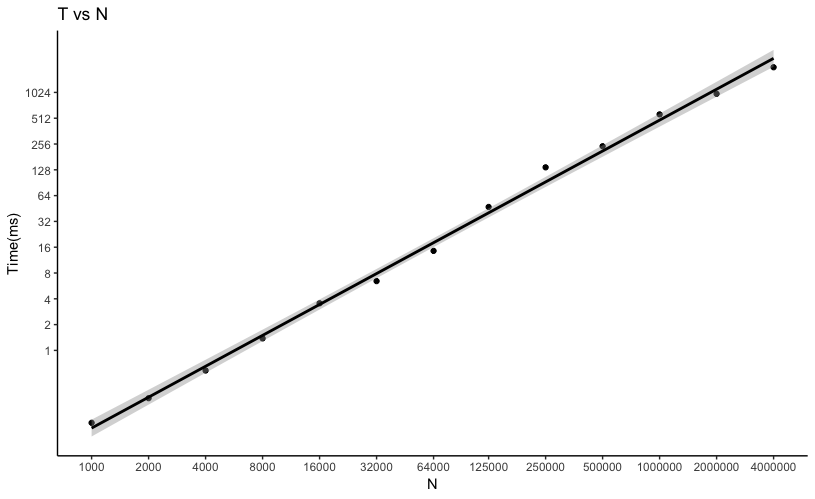}
\end{figure}

\item {}Comparison of Huskysort with system sort for numeric types (Integer, Double, Long, BigInteger, BigDecimal): Please refer to figure ~\ref{Fig:HS_BM_N} for results.
\begin{figure}[]
  
  \caption{Comparison of Huskysort with system sort (Numeric) }
  \Description{Comparison of multiple runs}
  \label{Fig:HS_BM_N}
  \centering
  \includegraphics[width=0.5\textwidth]{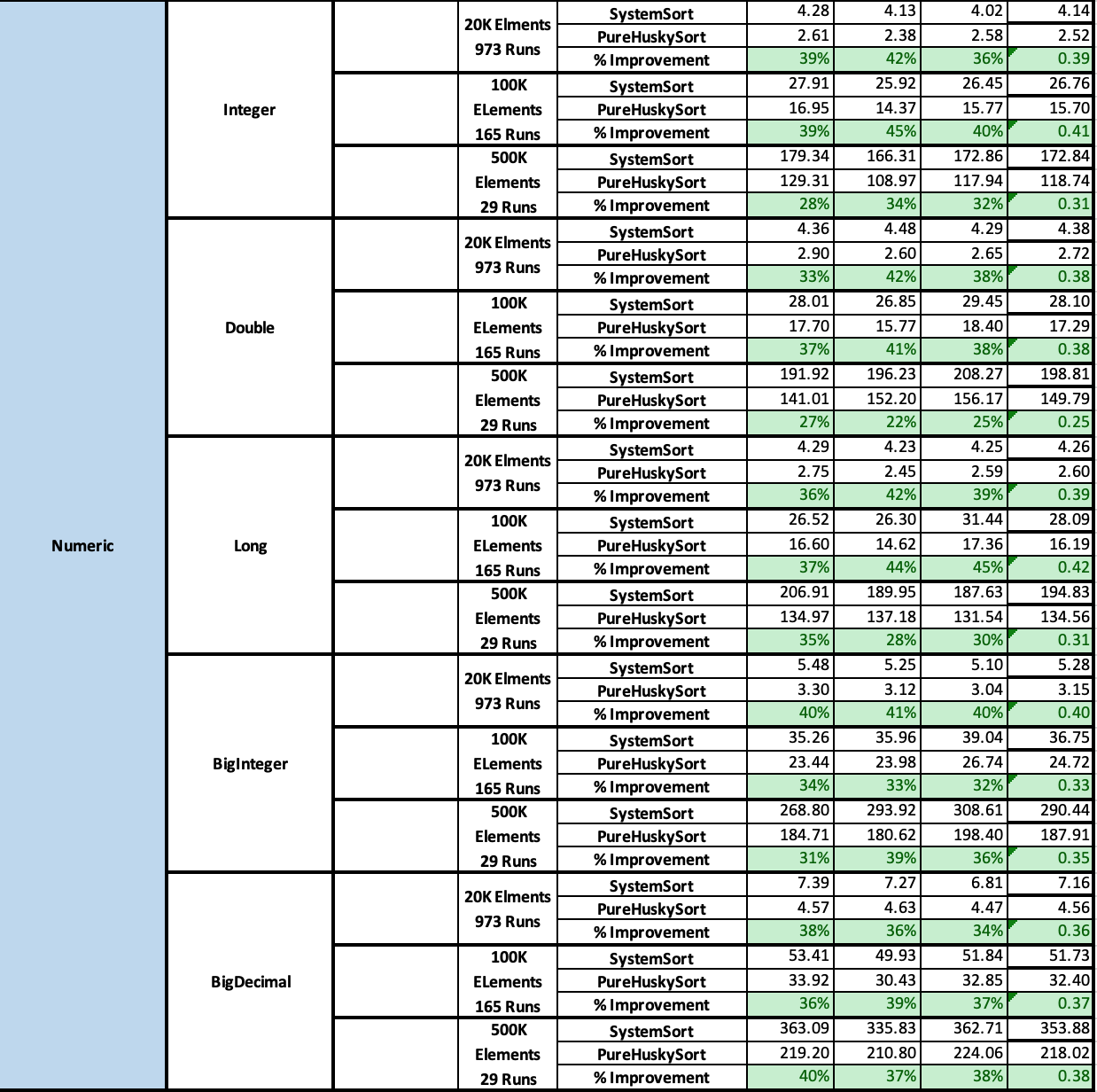}
\end{figure}

\item {}Comparison of Huskysort with system sort for (String types--English Words, Chinese Words): Please refer to figures ~\ref{Fig:HS_BM_SE}, ~\ref{Fig:HS_BM_SC} for results.
\begin{figure}[]
  
  \caption{Comparison of Huskysort with system sort (Strings--English Words) }
  \Description{Comparison of multiple runs}
  \label{Fig:HS_BM_SE}
  \centering
  \includegraphics[width=0.5\textwidth]{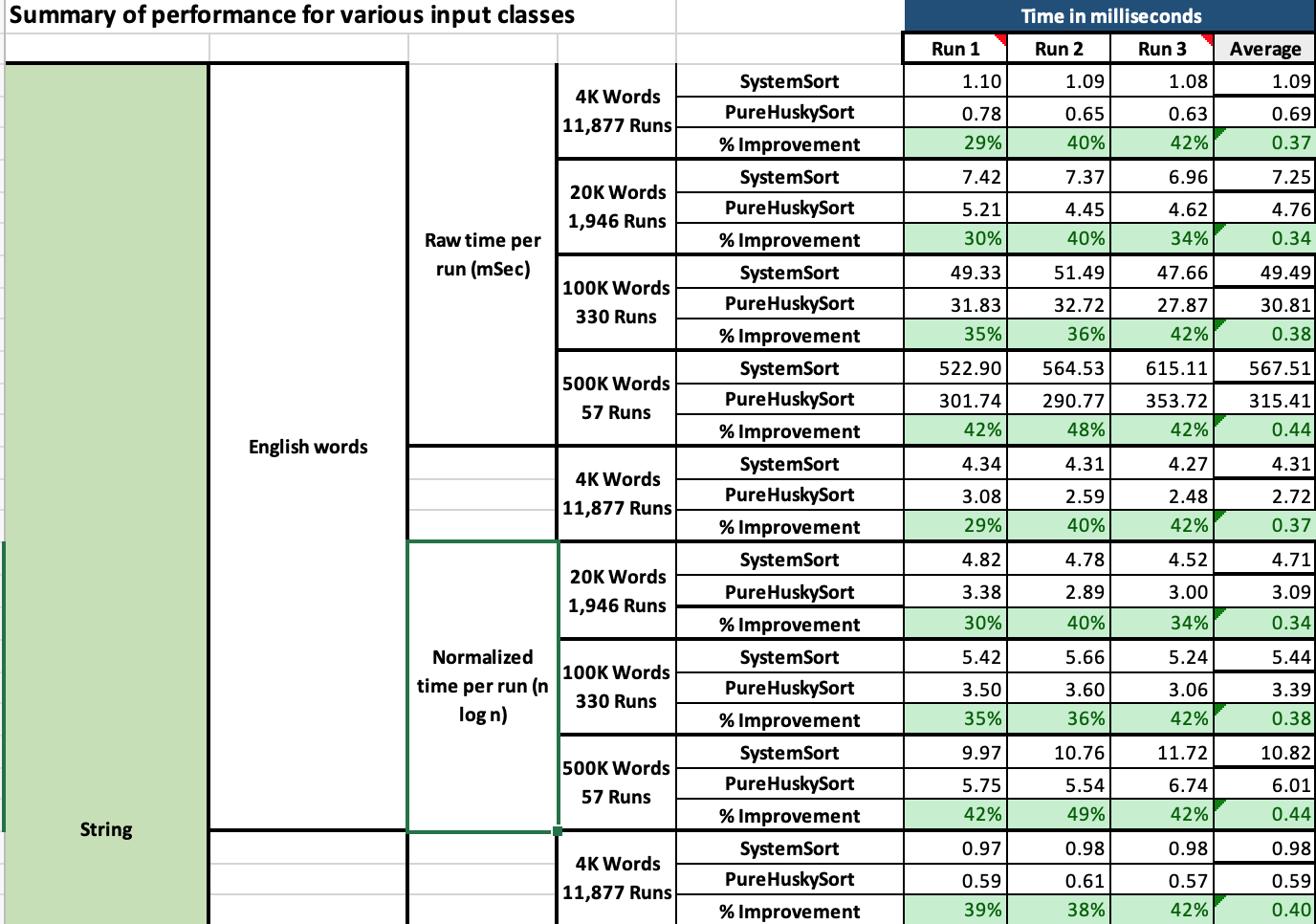}
\end{figure}
\begin{figure}[]
  
  \caption{Comparison of Huskysort with system sort (Strings--Chinese Words) }
  \Description{Comparison of multiple runs}
  \label{Fig:HS_BM_SC}
  \centering
  \includegraphics[width=0.5\textwidth]{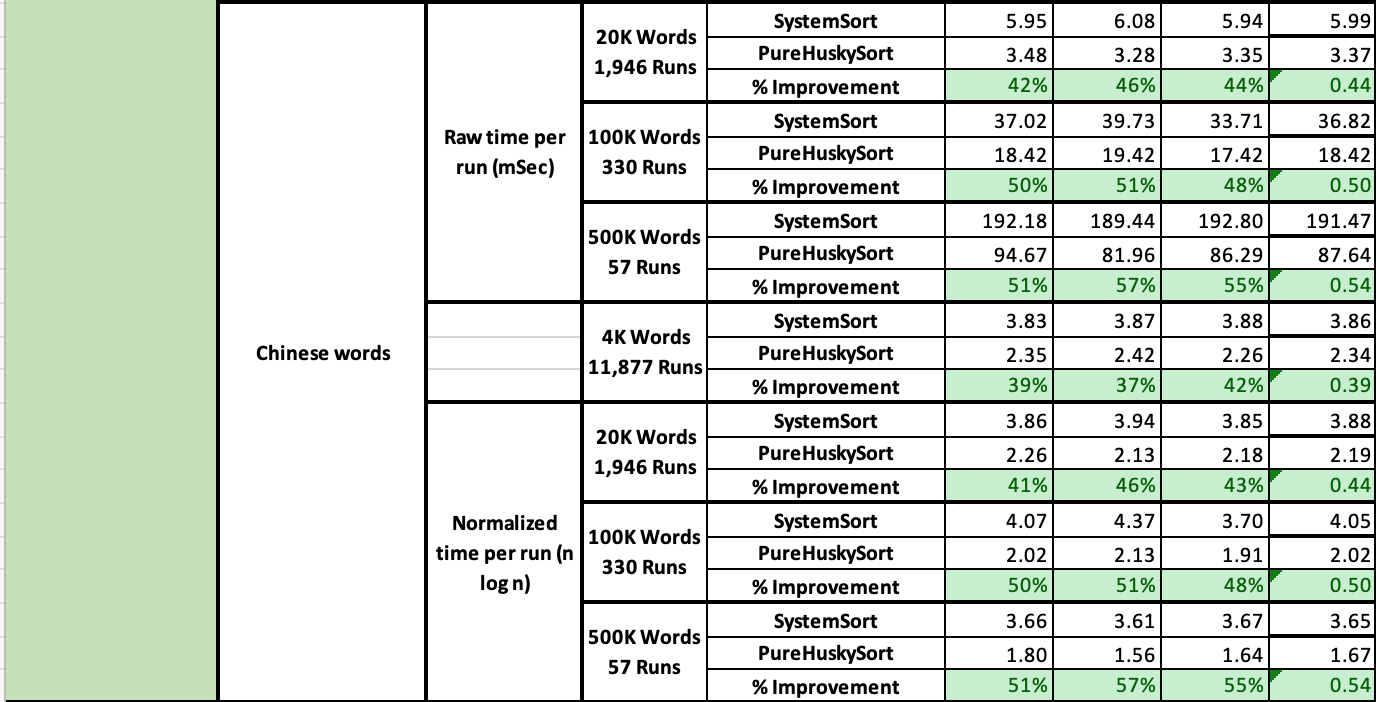}
\end{figure}

\item {}Comparison of Huskysort with system sort for Tuples: Please refer to figure ~\ref{Fig:HS_BM_T} for results.
\begin{figure}[]
  
  \caption{Comparison of Huskysort with system sort (Tuples) }
  \Description{Comparison of multiple runs}
  \label{Fig:HS_BM_T}
  \centering
  \includegraphics[width=0.5\textwidth]{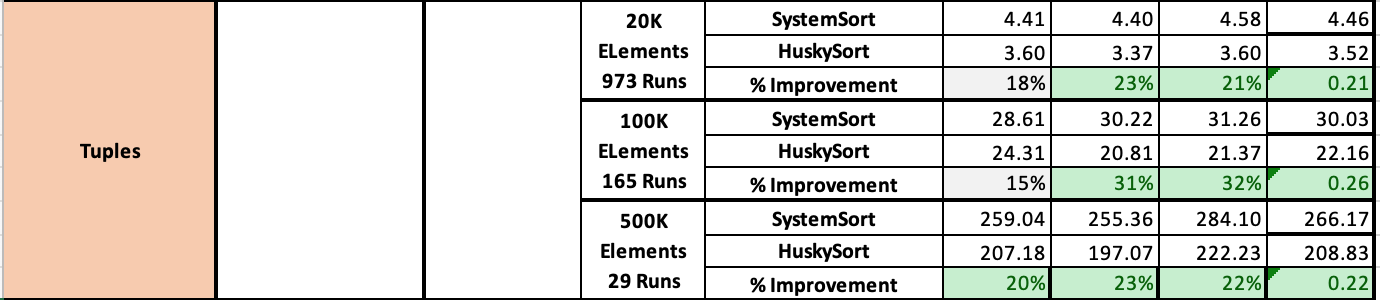}
\end{figure}

\end{itemize}

\subsection{Summary}
Huskysort’s target application is the sorting of randomly ordered arrays of objects on the JVM (Java Virtual Machine).
Let’s eliminate the non-use-cases first:
\begin{itemize}
\item {}Partially-ordered arrays: that’s to say, arrays with only a linear number of inversions as would be the case when adding a relatively small number of records to a large existing index.
For such use cases, Timsort (the Java system sort for objects) would be more suitable.

\item {}Arrays of primitives: in Java, a clear distinction exists between primitive types (char, byte, short, int, float, long, double) and objects.
Dual-pivot quicksort (the Java system sort for primitives) is already the fastest algorithm for sorting such types.
\end{itemize}
Thus, Huskysort is best for any object array where there’s no expectation of partial ordering.
Here is a summary of typical improvements in performance over Timsort:
Please refer to table ~\ref{tab:Improvements Summary}

\begin{table*}
  \caption{Improvements Summary}
  \label{tab:Improvements Summary}
  \begin{tabular}{ccl}
    \toprule
    Data Type  &  Percentage of Improvement\\
    \midrule
    \texttt{Strings of English words (4,000—500,000 elements)} &  33-41 \\
    \texttt{Strings of Chinese words (4,000—500,000 elements)}&  31-52\\
    \texttt{Tuples (20,000-500,000 elements))}&  14-19\\
    \texttt{Integers (20,000-500,000 elements))}&  31-41\\
    \texttt{Doubles (20,000-500,000 elements))}&  25-38\\
    \texttt{Longs (20,000-500,000 elements))}&  31-42\\
    \texttt{BigIntegers (20,000-500,000 elements))}& 35-40\\
    \texttt{BigDecimals (20,000-500,000 elements)}&  36-38\\
    \bottomrule
  \end{tabular}
\end{table*}

\section{Conclusion}

We have demonstrated that we can shift work from the linearithmic phase of sorting to the linear phase (both the prelude and the postlude) and effect an overall reduction in array accesses and thus processing time.
When sorting objects rather than primitives, Huskysort is always faster than dual-pivot quicksort.
It is faster than Timsort for arrays which are not partially ordered.
It is especially fast for Unicode character strings.

\begin{acks}
To Darshan Dedhia, Akshay Bhusare, Kartik Kumar for their help with this project.
\end{acks}

\bibliographystyle{ACM-Reference-Format}
\bibliography{sample-base}

\appendix
\section{Appendix}
\subsection{Discussion of coding accuracy}
It is not essential that coding is perfect for Huskysort to work well.
We have experimented with deliberately throwing off the encoding quite drastically.
If the probability of one of these deliberate mis-codings is less than approximately 25 percent, Huskysort can still save time.
The inference that we can take from this is that it is not critical that encodings are all very good.
We can be confident that we are still saving time up to about one in four errors.

These deliberate errors were tested for Bytes and Integers but not the other generic types.

\subsection{Handling of partially-sorted arrays}

Not surprisingly, Huskysort is best suited for sorting random arrays.
The benefits of partially-sorted arrays are enjoyed most by the merge sort family of algorithms (including Timsort).
However, we have studied the efficacy of a modified merge sort (it is required to manage both the objects and the longs) as the first pass of the Huskysort algorithm.
For random English String inputs, the merge-sort variation improves on system sort by approximately 20 percent whereas the normal intro-sort-based Huskysort improves on the system sort by more like 40 Percent.

However, it is not recommended to use Huskysort when arrays are partially ordered.
In such cases, the merge-sort variation of Huskysort will take anywhere between one and one half times and four times as long as Timsort.

\end{document}